\documentclass[12pt]{article}

\usepackage{amsmath,amssymb,amsfonts,amsthm,dsfont,dcolumn}
\usepackage[margin=-5pt]{caption}
\usepackage{graphicx,wrapfig}
\usepackage{tensor}
\usepackage{appendix}
\usepackage{physics}
\usepackage[font=small,skip=0pt]{caption}

\DeclareGraphicsExtensions{.pdf,.png,.jpg,.jpeg}

\newcolumntype{d}{D{.}{.}{0}}

\newlength{\textlength}
\newlength{\overlinelength}

\usepackage{color}

\def\e{\epsilon}
\def\m{\mu}
\def\n{\nu}
\def\a{\alpha}
\def\b{\beta}

\newcounter{subequation}[equation]

\def\be{\begin{equation}}
	\def\ee{\end{equation}}
\newcommand{\eel}[1]{\label{#1}\end{equation}}
\newcommand{\bea}{\begin{eqnarray}}
	\newcommand{\eea}{\end{eqnarray}}
\newcommand{\eeal}[1]{\label{#1}\end{eqnarray}}

\makeatletter

\def\thesubequation{\theequation\@alph\c@subequation}
\def\@subeqnnum{{\rm (\thesubequation)}}
\def\slabel#1{\@bsphack\if@filesw {\let\thepage\relax
		\xdef\@gtempa{\write\@auxout{\string
				\newlabel{#1}{{\thesubequation}{\thepage}}}}}\@gtempa
	\if@nobreak \ifvmode\nobreak\fi\fi\fi\@esphack}
\def\subeqnarray{\stepcounter{equation}
	\let\@currentlabel=\theequation\global\c@subequation\@ne
	\global\@eqnswtrue \global\@eqcnt\z@\tabskip\@centering\let\\=\@subeqncr
	
	$$\halign to \displaywidth\bgroup\@eqnsel\hskip\@centering
	$\displaystyle\tabskip\z@{##}$&\global\@eqcnt\@ne
	\hskip 2\arraycolsep \hfil${##}$\hfil
	&\global\@eqcnt\tw@ \hskip 2\arraycolsep
	$\displaystyle\tabskip\z@{##}$\hfil
	\tabskip\@centering&\llap{##}\tabskip\z@\cr}
\def\endsubeqnarray{\@@subeqncr\egroup
	$$\global\@ignoretrue}
\def\@subeqncr{{\ifnum0=?}\fi\@ifstar{\global\@eqpen\@M
		\@ysubeqncr}{\global\@eqpen\interdisplaylinepenalty \@ysubeqncr}}
\def\@ysubeqncr{\@ifnextchar [{\@xsubeqncr}{\@xsubeqncr[\z@]}}
\def\@xsubeqncr[#1]{\ifnum0=?{\fi}\@@subeqncr
	\noalign{\penalty\@eqpen\vskip\jot\vskip #1\relax}}
\def\@@subeqncr{\let\@tempa\relax
	\ifcase\@eqcnt \def\@tempa{& & &}\or \def\@tempa{& &}
	\else \def\@tempa{&}\fi
	\@tempa \if@eqnsw\@subeqnnum\refstepcounter{subequation}\fi
	\global\@eqnswtrue\global\@eqcnt\z@\cr}
\let\@ssubeqncr=\@subeqncr
\@namedef{subeqnarray*}{\def\@subeqncr{\nonumber\@ssubeqncr}\subeqnarray}

\@namedef{endsubeqnarray*}{\global\advance\c@equation\m@ne
	\nonumber\endsubeqnarray}

\makeatletter \@addtoreset{equation}{section} \makeatother
\renewcommand{\theequation}{\thesection.\arabic{equation}}

\newcount\hour
\newcount\minute
\newtoks\amorpm \hour=\time\divide\hour by 60\minute
=\time{\multiply\hour by 60 \global\advance\minute by-\hour}
\edef\standardtime{{\ifnum\hour<12 \global\amorpm={am}
		\else\global\amorpm={pm}\advance\hour by-12 \fi
		\ifnum\hour=0 \hour=12 \fi
		\number\hour:\ifnum\minute<10
		0\fi\number\minute\the\amorpm}}
\edef\militarytime{\number\hour:\ifnum\minute<10 0\fi\number\minute}

\def\draftlabel#1{{\@bsphack\if@filesw {\let\thepage\relax
			\xdef\@gtempa{\write\@auxout{\string
					\newlabel{#1}{{\@currentlabel}{\thepage}}}}}\@gtempa
		\if@nobreak \ifvmode\nobreak\fi\fi\fi\@esphack}
	\gdef\@eqnlabel{#1}}
\def\@eqnlabel{}
\def\@vacuum{}
\def\marginnote#1{}
\def\draftmarginnote#1{\marginpar{\raggedright\scriptsize\tt#1}}
\def\draft{
	\pagestyle{plain}
	\overfullrule=2pt
	\oddsidemargin -.5truein
	\def\@oddhead{\sl \phantom{\today\quad\militarytime} \hfil
		\smash{\Large\sl DRAFT} \hfil \today\quad\militarytime}
	\let\@evenhead\@oddhead
	\let\label=\draftlabel
	\let\marginnote=\draftmarginnote
	\def\ps@empty{\let\@mkboth\@gobbletwo
		\def\@oddfoot{\hfil \smash{\Large\sl DRAFT} \hfil}
		\let\@evenfoot\@oddhead}
	
	\def\@eqnnum{(\theequation)\rlap{\kern\marginparsep\tt\@eqnlabel}
		\global\let\@eqnlabel\@vacuum}  }

\renewcommand{\theequation}{\thesection.\arabic{equation}}

\def \bk {\bf{k}}

\def\e{\epsilon}

\def\m{\mu}
\def\n{\nu}
\def\a{\alpha}
\def\b{\beta}

\def \Tr {{\rm Tr}}
\def \ln {\mbox{ln}}

\def \m {\mu}
\def \n {\nu}

\def \NN{\mbox{\tiny NSNS}}

\def \N{\mbox{\tiny NS}}

\def\lg{\mbox{log}}

\begin{document}

	\begin{titlepage}
\hfill MCTP-16-11

		\begin{center}
\vskip 2 cm
			
			{\Large \bf Left-Right Entanglement Entropy of D$p$-branes}
			
			\vskip .7 cm

			\vskip 1 cm
			
			{\large   Leopoldo A. Pando Zayas$^1$, and Norma Quiroz$^2$, }
			
		\end{center}
		
		\centerline{\em ${}^1$ The Abdus Salam International Centre for Theoretical Physics}
		\centerline{\em Strada Costiera 11, 34014 Trieste, Italy}
		
		\vskip .4cm \centerline{\it ${}^1$ Michigan Center for Theoretical
			Physics}
		\centerline{ \it Randall Laboratory of Physics, The University of
			Michigan}
		\centerline{\it 450 Church Street, Ann Arbor, MI 48109-1120}
		
		\vskip .4cm \centerline{\it ${}^2$ Departamento de Ciencias Exactas }
		\centerline{\it Centro Universitario del Sur, Universidad de Guadalajara}
		\centerline{\it  Enrique Arreola Silva 883, C.P. 49000,  Cd. Guzm\'an, Jalisco, M\'exico}

\begin{abstract}
We compute the left-right entanglement entropy for Dp-branes  in string theory. We employ the CFT approach to string theory Dp-branes, in particular, its presentation as coherent states of the closed string sector. The entanglement entropy is computed as the von Neumann entropy for a density matrix resulting from integration over the left-moving degrees of freedom.  We discuss various crucial  ambiguities related to sums over spin structures and argue that different choices capture different physics; however, we advance a themodynamic argument that seems to favor a particular choice of replica. We also consider Dp branes on compact dimensions and verify that  the effects of T-duality act covariantly on the Dp brane entanglement entropy.  We find that generically the left-right entanglement entropy provides a suitable generalization of boundary entropy and of the D-brane tension.
		\end{abstract}
		
	\end{titlepage}

\section{Introduction}

String theory is the leading candidate for a theory of quantum gravity.  As such one of its most impressive results is a microscopic explanation for the Bekenstein-Hawking entropy of certain black holes \cite{Strominger:1996sh,Callan:1996dv}. In this context, and in further descriptions of black hole dynamics including low energy emission,  D-branes play a central role.

 Dp-branes can be viewed as coherent states in the corresponding string theory. In this manuscript, we explore some of its aspects from this point of view, in the hope of ultimately clarifying their role in the context of black holes.  The characteristic that we focus on is entanglement entropy. A natural motivation for the study of entanglement entropy in Dp-branes comes from the original considerations of entanglement entropy  and its potential relation to black hole entropy \cite{Bombelli:1986rw,Srednicki:1993im}.  More generally, the entanglement entropy of excited states and, particularly, of coherent states has been recently discussed in the literature. Various results have been obtained in the context of condensed matter  \cite{PhysRevB.88.075112}, including also applications to the entanglement of topologically ordered states \cite{2012PhRvL.108s6402Q}.

A natural entanglement quantity, the left-right entanglement entropy (LREE) of coherent states was discussed in \cite{PandoZayas:2014wsa}. Some interesting generalizations of the LREE include its formulation in the more general context of arbitrary rational CFT  \cite{Das:2015oha}, Ising model with interfaces \cite{Brehm:2015lja}, connection with level/rank duality \cite{Schnitzer:2015gpa}. More recent work \cite{Brehm:2015plf}  provided an alternative perspective on \cite{Das:2015oha}; other interesting directions were discussed in  \cite{Gutperle:2015kmw}.

Another important motivation for our work  comes from the hope of constructing a model similar to Maldacena's eternal black hole \cite{Maldacena:2001kr}. This model, albeit in higher dimensions, describe two coupled CFT's and focuses on the maximally entangled states.  Ultimately, we hope to connect the left-right entanglement entropy with the thermal field theory entropy which might have implications for a potential resolution to the  information paradox. There is, indeed, a precedent in the program of connecting entanglement entropy with thermal behavior and aspects of black hole physics as presented in \cite{Takayanagi:2010wp}.

In this manuscript, we take a step further in  the general considerations of LREE and address Dp-branes states in string theory. From a technical point of view,  a string theory is a CFT with added constraints. Since we work in the context of superstring theory we find various ambiguities in the choice and treatment of the various sectors.  In this paper, we provide a complete string-theoretic calculation of the left-right entanglement entropy of Dp branes.

	The rest of the paper is organized as follows. In section \ref{DpReview} we review the construction of $D$ branes as boundary states in the closed string sector. In section \ref{Sec:Density} we introduce details of the density matrix and our prescriptions for implementing the replica trick.   Section \ref{Sec:EE}  presents the left-right entanglement entropy of $Dp$ branes. Section \ref{Sec:UnrepNorm} considers the effect of the normalization while replicating the theory. In section \ref{Sec:Thermodynamics} we advance an argument, based on a potential thermodynamical interpretation, according to which a particular choice of correlations during the replica process is selected. We consider the case of compact spaces and the action of T-duality in section \ref{Sec:Compact}. We conclude and propose some open questions in section \ref{Sec:Conclusions}. We relegate to a series of appendices some technical material regarding properties of the theta functions and some explicit computations on the density matrix. 
	

\section{Dp Branes in String Theory}\label{DpReview}


In this section we briefly  review the relevant concepts of boundary states describing Dp-branes. A detailed  exposition of the subject is given in \cite{Gaberdiel:2000jr} from where we use most of our notation. The preponderant geometrical view of a D$p$-brane is that of a  hyperplane where open strings can end. In this context D$p$-branes  are boundaries for the end points of the string.
In the boundary conformal field theory approach, D-branes are boundary states of the closed string theory emmiting or absorbing closed string states. A boundary state is constructed by solving specific equations relating left- and right-modes of closed strings  at the boundary. We  work in the light-cone gauge with light-cone directions $x^0$ and $x^1$ which are taken to be Dirichlet directions.  In these conventions  $−1 \leq p \leq 7$. Therefore, the boundary states  actually  will be describing D-instantons.  However, one can perform an appropiate  Wick rotation to transform these states back into usual D-branes.
 For the transverse coordinates $x^\mu$, $\mu =  2, \dots, 9$ one can further split the index as $\mu = (\a, i)$ with $\a = 2, \ldots, p+2$   the Neumann directions and  $i =p+3, \ldots, 9$  the Dirichlet directions.

 The solution, using the notation  of \cite{Gaberdiel:1999ch}, is:

\be
\label{eq:bsmomentum}
\ket{\bk, \eta}_{\shortstack{$\scriptstyle{NSNS}$\\
		$\scriptstyle{RR}$}} =  \text{exp}\left\{\sum_{n>0}\left[
\frac{1}{n}\alpha^{\m}_{-n} S_{\m \n}\tilde{\alpha}^{\n}_{-n}
\right] + i\eta 	\sum_{r>0}\left[
\psi^{\m}_{-r}S_{\m \n}\tilde{\psi}^{\n}_{-r}
\right] \right\}
\ket{\bk,\eta}^0_{\shortstack{$\scriptstyle{NSNS}$\\
		$\scriptstyle{RR}$}}  \;,
\ee
where $\alpha_n$ denotes the bosonic operator with $n$ an integer and $\psi_r$ is the fermionic operator and depending on the sector, the mode number  $r$ is integer in R-R sector  or half-integer in NS-NS sector.  The matrix 
$S_{\m \n} = (\eta_{\a \b}, -\delta_{i,j})$ denotes the Neumann or Dirichlet boundary conditions of the D$p$-brane. The boundary ground state $\ket{\bk,\eta}^0_{\scriptstyle{NSNS}}$ is the unique ground state of the
NS-NS sector of the bulk theory and it carries momentum $\bk$ in the Dirichlet directions only. The definition of the boundary ground state in the  R-R sector is more involved but  as we will explain later, we will not be  interested in the R-R sector of the boundary states. For an explicit derivation of R-R ground state sector we refer the reader to \cite{Gaberdiel:2000jr}. The parameter $\eta = \pm$ denotes the two different spin structures that will play a prominent role in our story.

In order to  localize the boundary state  (\ref{eq:bsmomentum}) one has to take its Fourier transform:

\be
\label{eq:bslocal}
\ket{Bp, \bf{a}, \eta}_{\shortstack{$\scriptstyle{NSNS}$\\
	$\scriptstyle{RR}$}}
= {\cal N} \int \prod_{\n=0,1, p+3,\ldots,9} dk^\n e^{ik^\n a_\n}
\ket{\bk, \eta}_{\shortstack{$\scriptstyle{NSNS}$\\
		$\scriptstyle{RR}$}}  \;,
\ee
where  ${\bf a}$ denotes the position of the boundary state, ${\cal N}$ a normalization constant and the integration is over the Dirichlet (transverse) directions to the boundary.  In the following we consider the case ${\bf a}=0$ without loss of generality and in such case the boundary state is denoted simply as $\ket{Bp, \eta}$.

The boundary states \eqref{eq:bslocal} by themselves do not represent a Dp-brane. One has to further impose certain conditions based on physical grounds. One of the conditions required is GSO invariance of the boundary states, it ensures that the  boundary states couple only to the physical spectrum of the closed string theory.  
This condition implies that  physical boundary states are linear combinations of the  boundary states (\ref{eq:bslocal}). For the NS-NS sector, the GSO-invariant state is

\be
\label{eq:nsnsbrane}
\ket{Bp}_{\scriptstyle{NSNS}} = \sum_{\eta=\pm} \eta \ket{Bp,\eta}_{\scriptstyle{NSNS}} \,,
\ee
and the GSO invariant state for the R-R sector is
\be
\label{eq:rrbrane}
\ket{Bp}_{\scriptstyle{RR}} = \sum_{\eta=\pm}  \ket{Bp,\eta}_{\scriptstyle{RR}} \,.
\ee

Another important condition to be imposed  is the  open-closed duality where the tree-level amplitude of a boundary state with itself, describing the exchange of closed string states between them, has to be equivalent, after an appropiate conformal transformation, to the one-loop open string amplitude. From the CFT point of view, this is the Cardy condition. This requirement
 guarantees that the open  strings obtained by this duality have  consistent interactions with the original closed strings.
 
 This last condition allows to construct the D-branes as linear combinations of physical boundary states from different sectors. In Type IIA/B in ten dimensions there are two
 possibilities.  The BPS brane
\be
\label{eq:bps}
\ket{Dp} = \ket{Bp}_{NSNS}+\ket{Bp}_{RR}\,,
\ee
where $p$ is even (odd) in Type IIA (IIB). This is a stable brane and the open string spectrum of the open strings beginning and ending on two D-branes is supersymmetric.

There is also a  non-BPS brane defined as
\be
\label{eq:nonbps}
\ket{Dp}=  \ket{Bp}_{NSNS}\,,
\ee
for $p$ odd (even) in Type IIA (IIB). This is an unstable brane and the spectrum of open strings is never supersymmetric.

	\section{Density matrix for D-branes}\label{Sec:Density}

Given a system in a pure state $\ket{\psi}$, it can be characterized by the density operator $\rho = \ket{\psi}\bra{\psi}$ acting on the state space ${\cal H}$. Since the system is in a pure state, the density matrix is Hermitian and satisfy $\rho^2 = \rho$ and $\mbox{\Tr} \rho =1$.  If the system is factorized into  two different subsystems A and B with Hilbert space the tensor product ${\cal H}={\cal H}_A \otimes {\cal H}_B$, one can construct a reduced density matrix from $\rho$ describing the physical properties of one of the subsystems. For instance, the reduced density matrix of subsystem A  is obtained by taking the partial trace of $\rho$ with respect to subsystem B, that is,  $\rho_A={\mbox Tr}_B \rho$. At this point, $\rho_A$ allows one to describe the system A as if it were effectively isolated from system B. However,  systems A and B could be entangled showing quantum correlations between the two systems. There are several ways to measure entanglement. In this work we are interested in the entanglement entropy for subsystem A, which is defined as the von Neumann entropy  of the reduced density matrix $\rho_A$: 
\be
\label{eq:entropy}
S_A = -{\Tr}\,\rho_A\, {\lg}\, \rho_A,
\ee
a similar definition follows for $S_B$, moreover,  $S_A = S_B$ for a system in a  pure state. One efficient way to arrive at the entanglement entropy above is through the  R\'enyi entropies
$$
\label{eq:renyi}
S^{(n)}(\rho_A)= \frac{1}{1-n}\lg\,\mbox{Tr}\,\rho^n_{A} = S^{(n)}(\rho_B)\,,
$$
where $n \geq 0$. The entanglement entropy is then defined  as $S_A = \lim\limits_{n \rightarrow 1} S_A^{(n)}$. A convenient way to compute $\mbox{Tr}\,\rho^n_{A}$ for general real  $n$ is to employ the replica trick. This prescription consists in computing  $\Tr \rho^n_A$ for $n \in \mathbb{Z}^+$  and then taking the analytic continuation to $\mbox{Re}\, n >1$. By this approach one obtains
\be
\label{eq:trace}
\Tr \rho^n_A = \frac{Z_n(A)}{Z^n},
\ee
where $Z_n(A)$ is a partition function on a $n$-sheeted Riemann surface. 


Typically, subsystems $A$ and $B$ are defined as geometric regions of the whole system. In this paper, as in our previous one \cite{PandoZayas:2014wsa},  we will deviate from this geometric prescription. Our  whole system is an extended object in the supersymmetric closed type  IIA/B string theory. Excitations of the closed string can be decomposed into a superposition of modes that travel clockwise (left-moving) around the string and those that travel counter-clockwise (right-moving). Therefore in this work we define the subsystems A and B as the left- and  right-moving modes of  closed superstring theories, respectively. In particular, we are interested in computing entanglement entropy of left and right modes of boundary states defining Dp-branes in string theory.

\begin{figure}[h]
					\centering
					\includegraphics[scale=0.5]{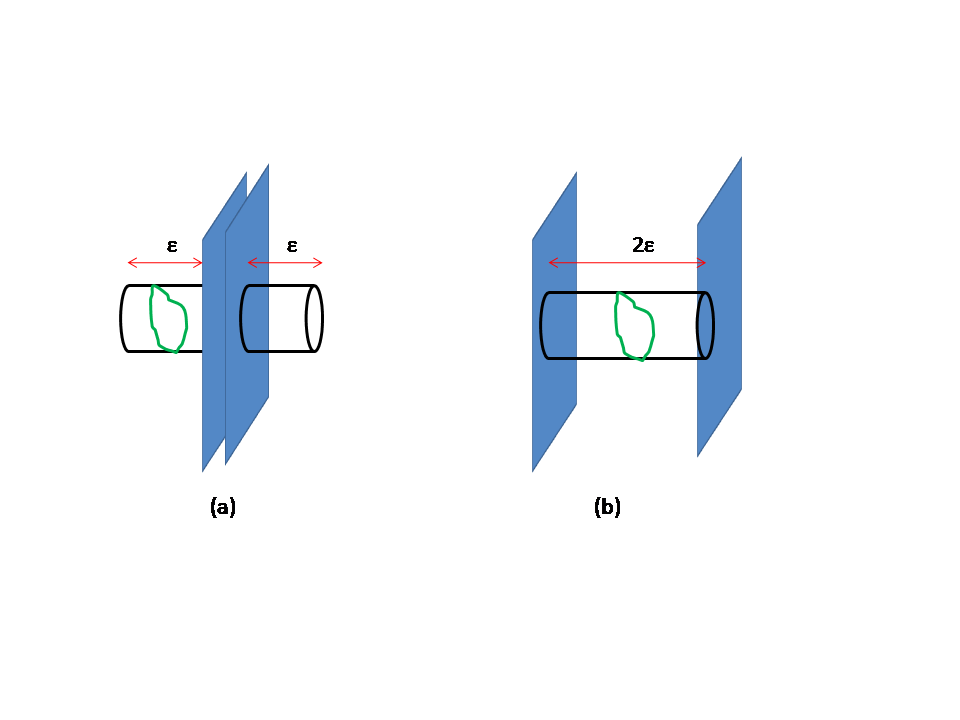}
					\caption{ The geometrical representation  of the  unnormalized $\rho_{\mbox{\tiny{NSNS}}}$ is given in (a). The lenght of each cylinder is $\epsilon$ with circumference 1. The figure (b) represents the interaction of two branes emitting and absorbing closed string states in a time $2 \epsilon$, the circumference of the cylinder is also 1. 					\label{fig:pf}}
				\end{figure}

	 Before discussing the reduced density matrix for the subsystem A we note that in order to define the density matrix for the branes we have to introduce a regularization  factor $e^{-\epsilon H}$ due to the fact that boundary states are non-normalizable. In this case $H$ is the Hamiltonian of the  supersymmetric closed string and the density matrix for the boundary state in the NS-NS sector or R-R sector is given as
	\be
	\label{eq:density}
	\rho^{\substack{\mbox{\tiny{NS}} \\\mbox{\tiny{R}}}}
	= \frac{1}{Z} e^{-\e H} \ket{Bp}_{\substack{\mbox{\tiny{NSNS}} \\
	\mbox{\tiny{RR}}}}\,{}_{\substack{\mbox{\tiny{NSNS}} \\\mbox{\tiny{RR}}}}\bra{Bp} e^{-\epsilon H}  \;,
	\ee
with $\ket{Bp}_{\substack{\mbox{\tiny{NSNS}} \\
		\mbox{\tiny{RR}}}}$ are defined by Eq. (\ref{eq:nsnsbrane}) and Eq. (\ref{eq:rrbrane}).  Note that one can consider the brane  to be localized in ${\bf a}$ as in eq.(\ref{eq:bslocal}), however, the term with the position of the brane does not contribute to trace of $\rho$ since the adjoint brane  is also taken to be localized at the same place. Therefore, we will consider  branes localized at the origin.
	For the BPS brane \eqref{eq:bps} the density operator is $\rho = \rho^{\mbox{\tiny{NS}}} \oplus \rho^{\mbox{\tiny{R}}}$.  A geometrical interpretation of $\rho$ is presented in Figure \ref{fig:pf} where  external closed string states propagate a time $\epsilon$, they are absorbed by a brane and after that, they are re-emitted by the same brane propagating  a time $\epsilon$. Note that the circumference of the cylinder displayed by the closed string state has been settled, to one, as is usual. 
	
Let us denote the closed string states as $\ket{{\it l}, \tilde{\it l},{\bf k}}$.
	 When taking the  matrix elements $\bra{{\bf k}, {\it l},\tilde{\it l}}\rho \ket{{\it l}', \tilde{\it l}',{\bf k}'}$   the trace imposes the periodic boundary conditions $\ket{{\it l}', \tilde{\it l}',{\bf k}'}=\ket{{\it l}, \tilde{\it l},{\bf k}}$. After taking the trace over the closed string states and impossing the condition  $\Tr \rho =1$ one finds that
	 
	 \be
	\label{eq:superpf}
	Z = {\cal N}^2 \Big( \frac{1}{2\epsilon}\Big)^{\frac{9-p}{2}}\frac{f_3^8(q)-f_4^8(q)-f_2^8(q)}{f_1^8(q)}
	= \bra{Dp}e^{-2 \epsilon H}\ket{Dp}\,.
	\ee
	
The result of tracing is given in the first equality, this term is the same as the  interaction force between two BPS-branes given
by the second equality.
	The geometry after taking the trace over  the closed strings is a cylinder with boundaries as in Fig. \ref{fig:pf}  representing the interaction between the branes exchanging closed string states   propagating a time $2\epsilon$ between the branes.
		The functions $f_i$ for $i=1,2,3,4$ are defined in appendix \ref{sec:fi} with $q=e^{-4\pi \epsilon}$ ($\tau = 4 i \epsilon$).  The partition function $Z$ so defined is identically zero (by Jacobi's  abstruse identity) revealing the fact that the exchange of NS-NS and R-R states cancel.  	Therefore, in this case the density matrix (\ref{eq:density}) is ill defined, since even after the regularization $e^{-\epsilon H}$ we are unable to normalize the correspondent density  matrix. 
	 Since we are interested in the entanglement between the left- and right-modes defining the brane, it is enough to take only the contribution of the boundary state in the NS-NS sector of the brane.  Therefore, we will take in our discussions only the NS-NS sector. In such case, the density matrix for the   BPS and non-BPS brane, is just
	  $\rho =\rho^{\mbox{\tiny{NS}}} $ and it is well  defined as we will see shortly.

	  In order to construct the reduced density matrix, we note that the space of states in the NS-NS sector of the closed superstring theory can be expressed as ${\cal H}_{\N}= {\cal H}^L_{\N} \otimes {\cal H}^R_{\N}$. An orthogonal basis  for the left (L) mode space is
	  \be
	  \label{eq:state}
	  \ket{l_L,{\bf k}} = \prod_{\mu=2}^{9}\prod_{n=0}^{\infty}\frac{1}{\sqrt{l_n !}} \Big(\frac{\alpha^{\mu_n}_{-n}}{\sqrt{n}}\Big)^{l_n} \prod_r \Big( \psi^{\mu_r}_{-r} \Big)^{l_r} \ket{0,{\bf k}}
	  \ee
	  with  $\ket{0,{\bf k}}$ the NS-ground state. A similar relation is given for the right (R) modes. Although the oscillator modes can be separated as left- and right modes, the closed string ground state carries only one momentum 
	 ${\bf k}$.

The boundary state given in \eqref{eq:bslocal} can be expressed in terms of the orthonormal basis of the space of states $\cal H_{\tiny{\mbox{NS}}}$. In fact, the boundary state is not a product state on such space but an entangled state in the right- and left-modes.	Then we can see the brane as a bipartite system consisting of subsystems A (right-modes) and B (left-modes).  To measure the entanglement  we define the reduced density matrix for system A. Since the NS-NS ground state  is labeled by the momentum ${\bf k}$ we have to sum also over the momenta. We define the reduced density matrix for A as:
	\be
	\label{eq:densityA}
	\rho_A = \Tr_B(\Tr_C\, \rho)\,,
	\ee
	where we have denoted the momentum by C. 
	Similarly $	\rho_B = \Tr_A(\Tr_C\, \rho)$   and $\rho_C = \Tr_A(\Tr_B\, \rho)$. For the bipartite entanglement, $S_A = S_B >0$
	but $S_C = 0$.
	
	 In order to simplify our notation we will drop  the labels of the states and wherever we write $\ket{Bp}$ and $\rho$ we mean $\ket{Bp}_{\mbox{\tiny{NSNS}}}$ and $\rho^{\mbox{\tiny{NS}}}$.
	The condition $\Tr \rho=1$ fixes the normalization constant as
	\be
	Z=\Tr \langle \langle \rho \rangle \rangle\,,
	\ee
	where we  use the notation $ \langle \langle \ldots \rangle \rangle$  to denote un-normalized density matrices  \cite{Gaberdiel:2000jr}.  The  trace is performed on ${\cal H}_{\N}$ and we obtain:
	
	\be
	\label{eq:z1}
	Z= 2 {\cal N}^2 \Bigl( \frac{1}{2\epsilon} \Bigr)^{\frac{9-p}{2}} \frac{f_3^8(q) - f_4^8(q)}{f^8_1(q)}.
	\ee
Note that this equation is similar to equation (\ref{eq:superpf}) except  for the contribution of the RR-sector given by $f_2(q)$ term. As in the supersymmetric case, this expression can be interpreted as a closed string state in the NS-NS sector, emitted from a NS-NS boundary state, the closed string state propagates for a time $2 \epsilon$ and then is absorbed by the same boundary state. Therefore, this normalization constant is the partition function describing the interaction of two  boundary states exchanging NS-NS closed string states
	\be
	\label{eq:partitionf}
	{Z}= \bra{Bp}e^{- 2\epsilon H}\ket{Bp}\,,
	\ee
	with $\ket{Bp}$ defined as in \eqref{eq:nsnsbrane}.

\section{Correlated and uncorrelated replica tricks}\label{Sec:EE}

\begin{figure}
	\centering
	\includegraphics[scale=0.5]{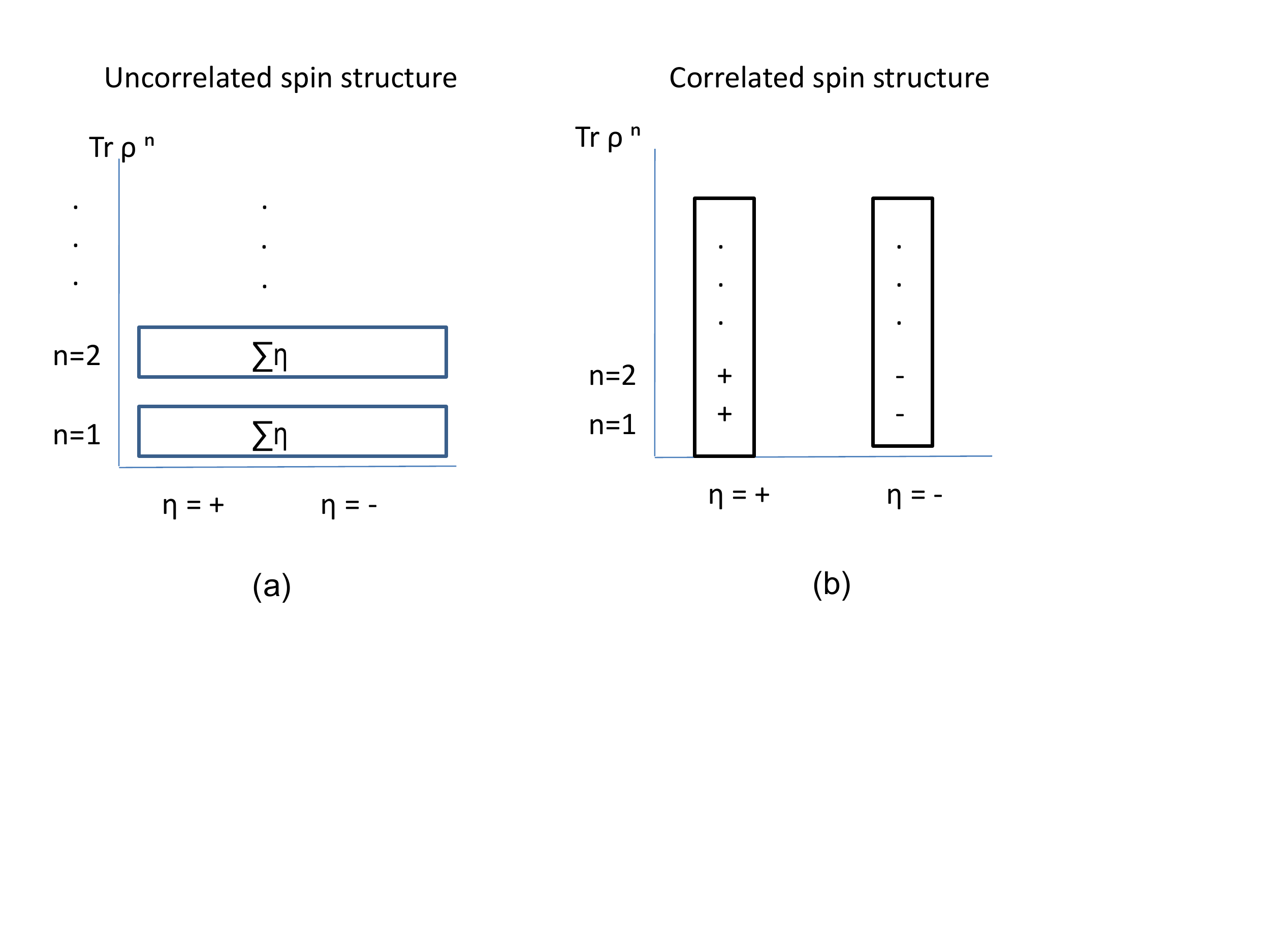}
		\caption{A graphic representation of the choices in replicating the theory as related to the treatment of spin structures. \label{fig:replicaspin}}
\end{figure}

We are interested in computing the replicated partition function $Z_n = \Tr \langle \langle \rho_A ^n \rangle \rangle$. When computing this trace, the right-mode oscillators satisfy the conditions ${\it l'}_i={\it l}_{i+1}$  and ${\it l}_1={\it l}'_{n}$ for $i=1,2,\ldots n$. A geometric representation of this process is given in Fig. \ref{fig:replicaspin} (for $n=3$).

	\begin{figure}
		\centering
		\includegraphics[scale=0.5]{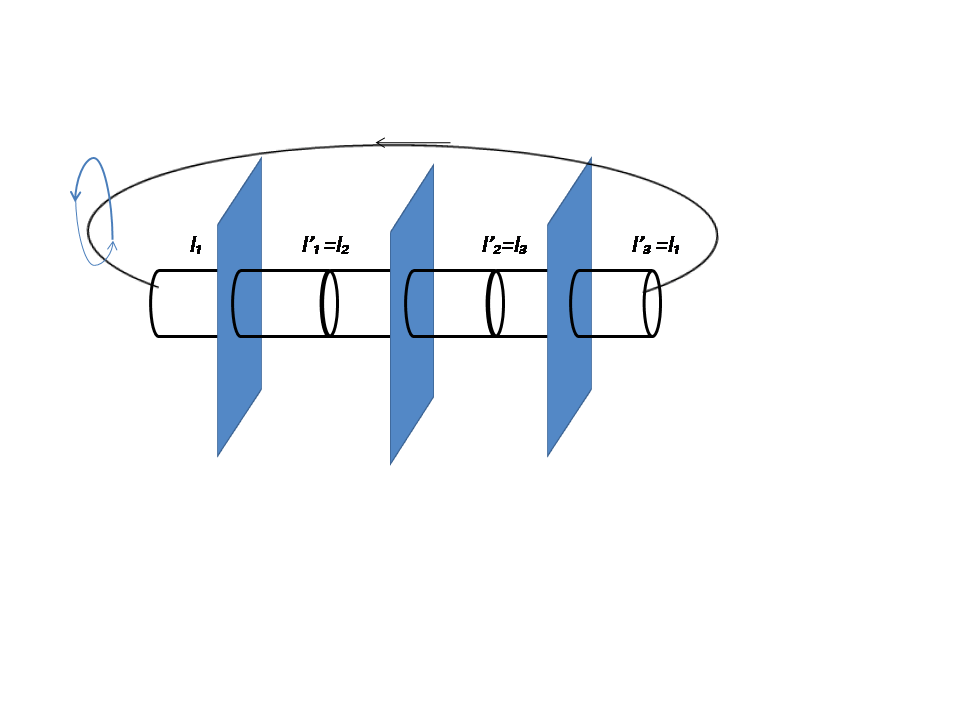}
		\caption{This graph represents a righ-mode closed string propagating between three different branes for $\Tr \rho_A^3$
				\label{fig:replica}}
	\end{figure}
Note that  our fermionic part of the reduced density matrix will contain a sum over the spin structures which is inherited from the fermionic part of  the boundary state.
 Here, we  consider two  possibilities to carry out the replica trick on the fermions according to \cite{Lokhande:2015zma}.  One can consider other possibilites but we will study only the ``extreme''  cases. In the first one, which is called the {\it uncorrelated spin structure}, the spin structures are summed over before the replication as is shown in Fig. \ref{fig:replicaspin} (a).  In this form, the spin structure of the copy denoted by  $n=1$ is disconnected with the spin structure of the next copy $n=2$ and so on.

The other extreme possibility for replicating is called the {\it correleted spin structure}. The spin structures in the fermionic part of the reduced density matrix are separated  and the replica is performed over each spin structure $\eta$ separately. After the replication, the spin structures are summed. This case is represented in Fig. \ref{fig:replicaspin}(b). 

In addition to the spin structures, the boundary states are also defined by the momentum. Due to the fact that one localizes the brane by the Fourier transform given in Eq. (\ref{eq:bsmomentum}), integration over momentum follows.  Therefore, we consider also two possibilities for replicating depending on whether one integrates over the momentum before of after  replicating. On the {\it uncorrelated momentum}, one integrates the momentum on each replica. For the {\it correlated momentum} one integrates over momentum after replicating.
	

\subsection{Correlated and uncorrelated spin structure with uncorrelated momenta}
Let us first describe the case of uncorrelated spin structure. The reduced density matrix ,
  $\langle \langle \rho_A  \rangle \rangle$ for system A as  given in Eq. \eqref{eq:trpA}. Note that the fermionic part of this unnormalized reduced density matrix is of the form
\be
\label{eq:rhof}
\langle \langle \rho_A  \rangle \rangle _f= 2 \sum_\eta \prod_{\mu=1}^{8}\prod_{r} \eta\Bigg[1 \otimes 1 + \eta\, q^{2r}\psi^{\mu}_{-r}\otimes \psi^{\mu}_r \Bigg]\ket{0}\otimes\bra{0}.
\ee
The factor of 2  comes from the fact that the product $\eta \eta'$ in  \eqref{eq:trpA} contributes twice to positive sign and twice to negative sign. 
When replicating over  the system A, we have to replicate this sum. In this way, at each step of the replica the spin structures have been summed  and, therefore,  they are uncorrelated between the $n$-copies of $\langle \langle \rho_A  \rangle \rangle$ as shown in Fig \ref{fig:replicaspin}(a).

 The partition function in this case is called the uncorrelated replica partition function, denoted by $Z^u_n$ and is given by

\be
\label{eq:zun}
Z^u_n= 2^{n-1} 2^n  {\cal N}^{2n} \Big(\frac{1}{2\epsilon}\Big)^{{(\frac{9-p}{2})n}}
\Bigg( \frac{f^8_3(q_n)-f^8_4(q_n)}{f^8_1(q_n)} \Bigg)\, ,
\ee
where $q_n=e^{i\pi  \tau_n}$ with $\tau_n = 4i\epsilon n$.  The factor $2^{n-1}$ results from the approach we are taking, since we are raising a sum over spin structures to a certain power. Note that as $\epsilon \rightarrow 0$,  we have $q_n \rightarrow 1$, therefore, in order to perform an expansion for small $\epsilon$ we have to make use of the  transformation $\tau_n \rightarrow -1/\tau_n$.   In this way $q_n = e^{-4\pi \epsilon n}$ transforms as  $\tilde q_n = e^{-\frac{\pi}{4\epsilon n}}$ and the transformation of the functions $f_i$ are given in Eq. (\ref{eq:mtf}). Therefore,

\be
 \frac{Z^u_n }{Z^n } \rightarrow  2^{n-1} \frac{(4\epsilon n)^4}{(4\epsilon)^{4n}}
\Bigg(\frac{f^8_1(\tilde q)}{f^8_1(\tilde q)-f^8_2(\tilde q)} \Bigg)^n \Bigg(\frac{f^8_3(\tilde q_n)-f^8_2(\tilde q_n)}{f^8_1(\tilde q_n)} \Bigg)\,.\ee
As $\epsilon \rightarrow 0$,  $\tilde q$  goes to zero and we can use Eq. (\ref{eq:fi}) to expand the functions $f_i$ for $i=1,2,3$   and get the leading terms
\be
\frac{Z^u_n }{Z^n } \rightarrow  2^{n-1}e^{-\frac{\pi}{4\epsilon}(n-\frac{1}{n})},
\ee
yielding an entanglement entropy of the form: 
\be
\label{sun}
S^u = \lim_{n\rightarrow 1} \frac{1}{1-n} \lg\,\Tr   \frac{Z^u_n}{Z^n} = - \mbox{ln}\, 2 + \frac{\pi}{2 \epsilon}.
\ee

Next we are going to analyze the correlated replica trick.  Starting from \eqref{eq:rhof} we separate the spin structures  and we write
\be
\langle \langle \rho_A  \rangle \rangle _f= 2 (\rho_+ - \rho_-),
\ee
where $\rho_{+/-}$  denotes the term in \eqref{eq:rhof}  with positive$/$negative spin structure.

We then take the replica trick for each spin structure as in Fig \ref{fig:replicaspin}(b). After that, we sum over the spin structures. The partition function obtained in this way is called the correlated partition function and is denoted as $Z_n^c$. It has the form

\be
\label{eq:zcn}
Z^c_n  = 2 {\cal N}^{2n} \Big(\frac{1}{2\epsilon}\Big)^{{(\frac{9-p}{2})n}}\Bigg( \frac{f^8_3(q_n)-f^8_4(q_n)}{f^8_1(q_n)}\Bigg).
\ee
Note that the factor of 2 in (\ref{eq:rhof}) is not replicated.
As in the previous  case, after the  transformation $\tau_n \rightarrow -1/\tau_n$ and considering the limit $\epsilon \rightarrow 0$  we get:

\be
\label{eq:scn}
S^c = \lim_{n\rightarrow 1} \frac{1}{1-n} \lg \,\Tr   \frac{Z^c_n}{Z^n} = \mbox{ln}\, 2 + \frac{\pi}{2 \epsilon}.
\ee

In both cases, the momentum was already summed before the replica and the replication was made  over system A only. Therefore, each replica has uncorrelated momenta.
Note that the normalization constant has been replicated in both cases.  Therefore,  the normalization constant as well as the  momentum contribution to $Z^u_n$ and $Z^c_n$ cancel with those terms in $Z^n$. 

\subsection{Uncorrelated spin structure with correlated momentum}

One could also consider the reduced density matrix  $\langle \langle \rho_{AC} \rangle \rangle = \Tr_B \rho$. This expression is given in Eq. (\ref{eq:trac}), where the trace is made over the left-modes only. We then perform the replication for this matrix.  Only after constructing such matrix we sum over the momentum.  In this case all the $n$-copies have the same momentum. This is what we call the correlated momentum.  In addition we have to consider the two forms to carry out the replica trick for the spin structures. 

We denote the replica partition function as  $Z'_n = \Tr \langle \langle \rho_{AC} ^n \rangle \rangle$.
When taking the correlated momenta and uncorrelated spin structure the partition function is:

\be
\label{eq:zpun}
Z'\,^u_n= 2^{n-1} 2^n  {\cal N}^{2n} \Big(\frac{1}{2\epsilon n}\Big)^{{(\frac{9-p}{2})}}
\Bigg( \frac{f^8_3(q_n)-f^8_4(q_n)}{f^8_1(q_n)} \Bigg)\,.
\ee
Note that the momentum contribution to this expression differs from that in \eqref{eq:z1}. Therefore, this term is not canceled with the momentum contribution of $Z^n$. In such case the entanglement entropy becomes
\be
\label{eq:spun}
S'\,^u = -\lg\, 2 + \frac{9-p}{2} \lg\, 2 + \frac{\pi}{2 \epsilon}.
\ee

For  correlated momentum an correlated spin structure over the replicas
\be
\label{eq:zpcn}
Z'\,^c_n= 2  {\cal N}^{2n} \Big(\frac{1}{2\epsilon n}\Big)^{{(\frac{9-p}{2})}}
\Bigg( \frac{f^8_3(q_n)-f^8_4(q_n)}{f^8_1(q_n)} \Bigg)\,.
\ee
The corresponding entanglement entropy becomes:
\be
\label{eq:spcn}
S'\,^c = \lg\, 2 + \frac{9-p}{2} \lg\, 2 + \frac{\pi}{2 \epsilon}.
\ee

We note that  all the previous replica partition function (\ref{eq:zcn}),(\ref{eq:zun}),(\ref{eq:zpun}) and (\ref{eq:zpcn}) reduce to (\ref{eq:z1}) for $n=1$:
\be
Z^c_{n=1} = Z^u_{n=1} = Z'\,^c_{n=1} = Z'\,^u_{n=1} = Z\,.
\ee

\section{Unreplicated normalization}\label{Sec:UnrepNorm}
The partition function $Z$ given in Eq.\eqref{eq:z1} is not invariant under the transformation $\tau \rightarrow -1/\tau$ since we are dealing with the NS-NS sector only. However, we noted that $Z$ can be expressed as $\bra{B}e^{-\epsilon H}\ket{B}$ which is the propagation amplitude between the same NS-NS boundary state. 
Such partition function satisfy, up to integrand factors the open/closed string duality. In \cite{Gaberdiel:2000jr} this duality has been described. For instance, for our NS-NS brane it was found that
\begin{align}
\int_0^\infty \, dl \bra{Bp,\eta} e^{-l H_c}\ket{Bp, \eta}& = {\cal N }^2 \int_0^\infty \, dl \Big(\frac{1}{l}\Big)^{\frac{9-p}{2}}\frac{f_3^8(q)}{f_1^8(q)}\\
&={\cal N}^2\, \frac{32 (2\pi)^{p+1}}{V_{p+1}} \int_0^\infty  \frac{dt}{2t} \Tr_{\N}\Big[e^{-tH_o}\Big],\\
\int_0^\infty \, dl \bra{Bp,\eta} e^{-l H_c}\ket{Bp, -\eta}& = {\cal N }^2 \int_0^\infty \, dl \Big(\frac{1}{l}\Big)^{\frac{9-p}{2}}\frac{f_3^8(q)}{f_1^8(q)}\\
& ={\cal N}^2\, \frac{32 (2\pi)^{p+1}}{V_{p+1}} \int_0^\infty \, \frac{dt}{2t} \Tr_{R}\Big[e^{-tH_o}\Big]\,,
\end{align}
where $l$ is the length of the cylinder in the closed string channel and $t$ the circumference  of the cylinder in the open string channel.
These relations fix the  normalization constant  for the non-BPS brane  \eqref{eq:nonbps}  as


\be
{\cal N}_{\mbox{\tiny{nonBPS}}}^2  =  \frac{V_{p+1}}{64 (2 \pi)^{p+1}},
\ee
and for the BPS brane \eqref{eq:bps} the NS-NS normalization constant as
\be
{\cal N}_{\mbox{\tiny{BPS}}}^2 = \frac{V_{p+1}}{128 (2 \pi)^{p+1}}\,.
\ee

 We can ask then, if some of the replicated partition functions $Z_n$ presented in Sec. 3 satisfy  the closed-open string duality. In all cases above, we have considered the natural way of computing the trace, taking the  replication of the normalization constant giving rise to the factor ${\cal N}^{2n}$, therefore, none of the replicated partition functions (\ref{eq:zun}), (\ref{eq:zcn}),(\ref{eq:zpun}) and (\ref{eq:zpcn}) satisfy this duality.  However,  we can take another approach for the replication $Z_n$. Following \cite{Lokhande:2015zma} we redefine $Z_n$ multiplying it by a term independent of $\epsilon$, that is: 
 
 \be
 \tilde Z_n = {\cal N}^{2(1-n)}Z_n\,.
 \ee
This is equivalent to requiring the normalization constant ${\cal N}$ not to be  replicated. From all the replicated partition function $Z_n$, correlated and uncorrelated, only $Z'{}_n^c$ with the redefinition given above, can be related to the closed-open string duality. Then, the partition function with uncorrelated normalization and correlated momentum and spin structure has the form:
\be
\tilde Z'\,^c_n =  2  {\cal N}^{2} \Big(\frac{1}{2\epsilon n}\Big)^{{(\frac{9-p}{2})}}
\Bigg( \frac{f^8_3(q_n)-f^8_4(q_n)}{f^8_1(q_n)} \Bigg)\,,
\ee
which can be expressed as
\be
\label{eq:pcupf}
\tilde Z'\,^c_n= \bra{Bp}e^{-2\epsilon n H}\ket{Bp}\,.
\ee
 This represents the  amplitude for NS-NS closed string states  propagating  between the same brane. Such expression satisfies the open-closed duality under which the closed string cylinder of circunderence 1 and lengnth $2 \epsilon n$ is changed to a cylinder in the open string channel with lenght 1 and circunfernce  $1/4\epsilon n$.
 The normalization constant does not depend on the geometry of the cylinder.  In the limit $\epsilon \rightarrow 0$, the leading terms in the trace are given by the lightest open string states. 
In this case we have:

\be
\label{eq:phystrace}
\frac{{\tilde Z}'^c_n}{Z^n} = 2^{1-n}{\cal N}^{2(1-n)} \Big( \frac{1}{2\epsilon n} \Big)^{\frac{9-p}{2}}
\Big(\frac{1}{2\epsilon } \Big)^{n\frac{p-9}{2}}
\Bigg(\frac{f^8_1(\tilde q)}{f^8_3(\tilde q)-f^8_2(\tilde q)} \Bigg)^n \Bigg(\frac{f^8_3(\tilde q_n)-f^8_2(\tilde q_n)}{f^8_1(\tilde q_n)} \Bigg)\frac{(4\epsilon n)^4}{(4\epsilon)^{4n} }\,.
\ee

 The entanglement entropy for the non-BPS brane \eqref{eq:nonbps} is then
\be
\label{eq:beet}
\tilde S'\,^c = \lg\,2 + \lg \frac{V_{p+1}}{64(2\pi)^{p+1}} + \frac{9-p}{2} \lg\,2  + \frac{\pi}{2\epsilon}\,.
\ee
We note that this expression depends on the normalization of the brane (related to the tension of the brane). In \cite{Harvey:1999gq} 
an expression for the tension and the boundary entropy was found. Here, we observe that the term of logaritm of the volumen is precisely the boundary entropy of the brane, given as $\bra{0}\ket{Bp}$ \cite{Affleck:1991tk}.  In this way we found a relation between the  left-right entanglement entropy and the boundary entropy of the brane. 
 After simplifications \eqref{eq:beet} reduces to
\be
\tilde S'\,^c  = \lg \frac{V_{p+1}}{(2\pi)^{p+1}} - \frac{p+1}{2}\lg\, 2 + \frac{\pi}{2\epsilon}.
\ee
Similarly, for the BPS brane, the normalization constant is one half the normalization of the non-BPS brane due to  the GSO projection, therefore the entanglement entropy is
\be
\tilde S'\,^c_n=  \lg \frac{V_{p+1}}{2(2\pi)^{p+1}}- \frac{p+1}{2}\lg\, 2  + \frac{\pi}{2\epsilon}\,.
\ee

\section{A potential thermodynamic  interpretation} \label{Sec:Thermodynamics}

Given the ambiguities that permeate the definition of entanglement entropy, it makes sense to look for additional criteria that might help us choose a particular prescription over others. With this aim we would like to  consider a structure akin to a thermodynamic limit. So far, the string theory and the boundary states have been defined at zero temperature. However, we can try to relate (in a fictitious way) the factor $2\epsilon$  in \eqref{eq:pcupf}  with the temperature. In such case, the high-temperature limit of equation \eqref{eq:phystrace} is dominated by the lowest energy states, that is, the  massless closed string modes:
	$$
	\Tr 
	\frac{{\tilde Z}'^c_n}{Z^n}\sim {\cal N}^{2(1-n)} \Big( \frac{1}{2\epsilon n} \Big)^{\frac{9-p}{2}}
	\Big(\frac{1}{2\epsilon } \Big)^{n\frac{p-9}{2}}
	16^{1-n}\,.
	$$ 
 For this case the left-right entanglement entropy is:
 \be
 S_A (\beta) = \lim\limits_{n \rightarrow 1}\frac{1}{1-n} \Tr\frac{{\tilde Z}'^c_n}{Z^n}= 5 \lg\, 2 + \lg\, {\cal N}^2 + \Bigl(\frac{9-p}{2}\Bigr) \lg\, \beta  - \frac{9-p}{2}.
  \ee
 Our key observation is that this result agrees with the thermodynamical entropy computed from the partition function \eqref{eq:z1} in the limit $\beta = 2\epsilon \rightarrow 0$
\be
S_{th}= \beta^2 \frac{\partial}{\partial \beta} \Biggl(-\frac{1}{\beta}\lg\,Z  \Biggr) = S_A(\beta).
\ee
A number of comments are in order.  First, according to this criterion, it seems that the  uncorrelated normalization and correlated momentum and spin structure entanglement entropy is the one that is favored. We highlight, however, that this criterion is just that, a way to concoct an physically reasonable extra condition that might help us understand which of the various procedures (correlated or uncorrelated replicas) leads to a more meaningful result.   

Second, the high temperature limit is, in a sense a fairly universal limit. Indeed, one expects that at very high temperatures the system will be guided by universal ideas of equipartition of energy. The crucial point here is that, due to the nature of string theory, this high temperature limit essentially picks out the massless states of the theory.  This is quite different from the standard particle picture where it is the low temperature limit that picks the states with the lowest energies. 

Perhaps more speculatively, it is worth mentioning that there have been some other contexts where the entanglement entropy seems to be connected to a thermodynamic entropy as evidenced in first-law like relations discussed in 
\cite{Blanco:2013joa,Wong:2013gua}. Such thermodynamic properties of the entanglement entropy were, in a sense, anticipated in particular cases like \cite{Klebanov:2007ws} where by considering varying the length of the entanglement region a confinement/deconfiment transition was suggested. This reasoning leads to the treatment, effectively of the length of the interval as an effective temperature. More importantly, this effective temperature, as discussed in, for example \cite{Faraggi:2007fu}, is independent of the ``temperature'' of the dual supergravity background.

	\section{Compactification and T-duality}\label{Sec:Compact}
	
For the case of compactification on a torus $T^k$, the boundary states describing D-branes have some small  differences from the uncompatifed case. The part corresponding to the non-zero modes is the same as before but the part of the zero-modes is modified since they are now characterized by  quantized momenta and winding number along the compact directions.  Along the Dirichlet directions the boundary state  carries momentum $\frac{m_i}{R_i}$ and is denoted by $\ket{B,m_i}$. The state $\ket{B,w_i}$ is the boundary state along the Newman directions carrying winding number $w_iR$.
 In the localized boundary state the integral of momenta over compact directions is replaced by a sum over the momenta $m_i$. One also has to sum over the winding numbers.
 For this case, there are $r+1$ Neumann directions along the non-compactified coordinates and $s$ Neumann compactified  directions, with $p=r+s$.  The boundary state is:

 \be
 \label{eq:compact}
 \ket{Bp,{\bf a},\theta, \eta} = {\cal N}^2 \int \prod_{\mu} dp^{\mu}e^{ip\cdot a}\prod_{i=1}^s \sum_{w_i}e^{i\theta \cdot w R}\prod_{j=1}^{k-s}\sum_{m_j}e^{i{\hat a}\cdot \frac{m}{R}}\ket{Bp,{\bf p},{\bf m}, {\bf w}, \eta}\,,
 \ee
where ${\bf a}=(a_\mu,\hat a_j)$ denotes the position of the brane along the transverse non-compact and compact directions to the brane, respectively. The parameter $\theta$ is a Wilson line associated with the $U(1)$ gauge field living on the Dp-brane.

As in the previous case $Z= \Tr \langle \langle \rho_A ^n \rangle \rangle$ is the amplitude between two coincident branes $\bra{Bp,\theta,\eta}e^{-\epsilon H}\ket{Bp,\theta,\eta}$:
\be
Z= 2 {\cal N}^2 \Big(\frac{1}{2\epsilon} \Big)^{\frac{9-(r+k)}{2}}\prod_{i=1}^s \sum_{w_j}
e^{-2\pi \epsilon (w_i R_i)^2}
\prod_{j=1}^{k-s} \sum_{m_j}
e^{-2\pi \epsilon (\frac{m_j}{ R_j})^2}
\frac{f_3^8(q) - f_4^8(q)}{f^8_1(q)}\,.
\ee
Following the prescription stated above we find an interesting case, the replicated partition function $Z'{}^c_n$ is constructed by taking the following combination: uncorrelated normalization  and uncorrelated momenta along $T^n$, correlated momentum on the non-compact directions and correlated spin structures over the replicas. The result is 

\be
\label{eq:compactphysical}
Z'\,^c_n= 2  {\cal N}^{2} \Big(\frac{1}{2\epsilon  n}\Big)^{{\frac{9-(r+k)}{2}}}
\Bigg(
\prod_{i=1}^s \sum_{w_j}
e^{-2\pi \epsilon (w_i R_i)^2}
\Bigg)^n
\Bigg(\prod_{j=1}^{k-s} \sum_{m_j}
e^{-2\pi \epsilon (\frac{m_j}{ R_j})^2}
\Bigg)^n
\Bigg( \frac{f^8_3(q_n)-f^8_4(q_n)}{f^8_1(q_n)} \Bigg)\,.
\ee

The corresponding entanglement entropy  is:

\be
\tilde S'^c = \lg\, 2 + \lg {\cal N}^2 + \frac{9-r}{2}\lg\, 2 + \frac{\pi}{2\epsilon}.
\ee
For the non-BPS brane the normalization constant is 
$${\cal N}^2 = \frac{V_{r+1}}{64(2\pi)^{r+1}}\frac{\prod_{j=1}^{k-s}R_j }{\prod_{i=1}^s R_i} \,. $$

Therefore, the entanglement entropy  for a Non-BPS D$p$-brane   on $T^k$ with $r$ non-compact  and  $s$ compact Neumann  directions is:
\be
\label{eq:compacentropy}
S'^c(r,s) = \lg\, 2 + \lg\frac{V_{r+1}}{64(2\pi)^{r+1}} + \lg \prod_{j=1}^{k-s}R_j +\lg\prod_{i=1}^{s} \frac{1}{R_i}+ \frac{9-r}{2}\lg\, 2 + \frac{\pi}{2\epsilon}.
\ee

Note that T-duality interchanges Neumann  and Dirichlet directions.  In particular, the radius of compactification is changed as $R \rightarrow \frac{1}{R}$. Starting form \eqref{eq:compacentropy} one obtains the entanglement entropy $S'^c(r,s-1)$ of a D($p-1$) brane performing a T-dualtiy on one of the compact Neumann directions. Similarly, the entanglement entropy $S'^c(r,s+1)$ of a  D($p+1$)-brane is obtained by T-duality on one of the compactified Dirichlet directions.
We note that $S^c(r,0)$ describes the  entanglement entropy with all compact directions Dirichlet. $S^c(r,k)$  is the entanglement entropy with all compact Neumann  directions.  Then, one can verify that 
\be
S^c(r,0) > S^c(r,k).
\ee

It is worth pointing out that our main result in Eq. (\ref{eq:compacentropy}) is compatible with the analysis of \cite{Elitzur:1998va} which considered the boundary entropy of various compactified  configurations, albeit in the bosonic string. In particular, as pointed out in \cite{Elitzur:1998va} in the context of toroidal compactification, there is a path from the results in terms of boundary entropy to some of the spacetime moduli. The key observation is to view the radii entering in Eq. (\ref{eq:compacentropy}) as elements of the metric, further introduce a B-field and construct the complex moduli such as $\rho=B_{12}+i\sqrt{\det G}$ on which the entropy will depend. We will not pursue this direction further in this paper but it is clearly interesting and forms part of the  more ambitious goal of translating worldsheet left-right entanglement entropy computations into spacetime statements.


\section{Conclusions}\label{Sec:Conclusions}
	
We have computed the left-right entanglement entropy of string theory Dp branes. Our results can be interpreted as a generalization of the tension of a Dp brane as follows from \cite{Harvey:1999gq}. This is not surprising, indeed, our previous work in \cite{PandoZayas:2014wsa} has shown that  the left-right entanglement entropy  includes the boundary entropy  which is precisely related to the tension of a D-brane including the more general case of compactifications  \cite{Harvey:1999gq,Elitzur:1998va}. The new aspect which we have addressed in the present manuscript is the presence, in the case of the superstrings Dp branes, of various sectors.

It is clarifying to contrast  our computation  with that of the string tension given by Polchinski \cite{Polchinski:1995mt}. It has been important for us to consider only a given spin structure and we have chosen the NS one. We have verified that the entanglement entropy vanishes if one includes all the spin structures, just as the overall cylinder diagram vanishes in the tension computation of Polchinski. It is interesting that our computation seems to indicate that the type of entanglement between the left and right sectors depends, indeed, on the type of interaction that is allowed. Namely, as we saw in the text, considering the NS and the RR sectors together leads, due to supersymmetry, to a cancellation. We have argued that our computation is conceptually analogous to \cite{Harvey:1999gq} which, for the purpose of the D-brane tension, focused exclusively on the interaction with the graviton which resides in the NS-NS sector.  Again, choosing  a particular spin structure on the world sheet is equivalent to choosing a particular type of spacetime interactions and can thus  quantify how entangled two subsystems are due to a particular type of interaction. This is particular of superstring theory where we see clearly the contributions of the gravitational (NS) sector and of the charged (RR) sector.		

We have also considered Dp branes on compactified spaces and subsequently studied the left-right entanglement entropy under T-duality. The resulting expression, Eq. (\ref{eq:compacentropy} ), is covariant under T-duality that sends a $Dp$ into a $D(p\pm 1)$. 

In this manuscript we have  presented the completely un-correlated and completely correlated entanglement entropies. As stated clearly in the conclusions of \cite{Lokhande:2015zma}, there is a set of partitions functions in which one can choose to partially  correlate the spin structures. It would be important to understand, perhaps based on some other physical criterion, which of the partitions functions is more appropriate. Recall that in the case reported in \cite{Lokhande:2015zma} it was found that, depending of the size of the interval for the entanglement entropy, one was instructed to take the uncorrelated version for $(\ell \to 0)$ while the correlated, taking the product over replicas before summing over spin structures when the interval was close to the whole system size $(\ell \to L)$. Along these lines, we have suggested that the entanglement entropy that most naturally accommodates a thermodynamical interpretation is the one computed using uncorrelated normalization and correlated momentum and spin structure.  This is a topic that clearly deserves more attention.

One of the directions that is left somehow unexplored in this manuscript but that constitutes a driving motivation is the interpretation of our worldsheet results in terms of spacetime. The fundamental prototype is the interesting computation of Di Vecchia and collaborators 		\cite{DiVecchia:1997pr} who intuited spacetime properties of D-branes by using worldsheet computations. The p-dependence of the left-right entanglement entropy of a Dp brane in formulas such as Eq. (\ref{eq:compacentropy} )  seems to be  a first incipient step in this direction. 
		
Another interesting direction to explore would be to understand to what extend the entropy computed in this manuscript respects a first-law like relation for the entanglement entropy and the energy of excited states. Such relations have been obtained purely in field theoretic terms \cite{Blanco:2013joa}, \cite{Wong:2013gua} but  in rather symmetry restricted setups. Equivalently, it would be interesting to understand the structure of the modular Hamiltonian underlying our computation of the reduced density matrix. A connection to a first-law like relation might also lead  to a physical criterion as the one used in \cite{Lokhande:2015zma} where the entanglement entropy was expected to be related in a particular way to the thermodynamic entropy as  suggested in \cite{Azeyanagi:2007bj} and later proven in \cite{Chen:2014ehg}.

More speculatively, it would be interesting to explore whether this D-brane worldsheet computation can be used  to understand certain field theory entanglement entropy aspects. For example there might be aspects of ${\cal N}=4$ SYM that could be obtained as a particular limit of these configurations of open strings. We hope to explore some of these interesting issues in the future.

Finally, and perhaps more immediately, it would be interesting to place our explicit computations in the more general context of RCFT with supersymmetry. This effort will certainly build upon the beautiful analysis presented in \cite{Das:2015oha} and the background provided in \cite{Harvey:1999gq}. We hope to soon  report some progress in this direction.

\section*{Acknowledgments}

We are very thankful to  Sunil Mukhi, Cornelius Schmidt-Colinet, and \'Alvaro V\'eliz-Osorio for reading a preliminary version of this manuscript and important comments. L. Pando Zayas acknowledges the hospitality of CRM, Montreal where important conversations took place with Gautam Mandal about the direction of this project.

\appendix
\section{The $f_i$ functions}
\label{sec:fi}
We follow  \cite{Gaberdiel:1999ch} to define the functions $f_i$ in terms of the Jacobi Theta functions as: 
\begin{align}
\label{eq:fi}
f_1(q) &= q^{\frac{1}{12}}\prod_{m=1}^{\infty}(1-q^{2m})= 
(2\pi)^{-1/3}\theta'_1(0|\tau)^{1/3}\nonumber\\
f_2(q) &= q^{\frac{1}{12}} \prod_{m=1}^{\infty}(1+q^{2m})=  (2\pi)^{1/6}\theta_2(0|\tau)^{1/2}\theta'_1(0|\tau)^{-1/6}\nonumber\\
f_3(q)& = q^{-\frac{1}{24}}\prod_{m=1}^{\infty}(1+q^{2m-1})= (2\pi)^{1/6}\theta_3(0|\tau)^{1/2}\theta'_1(0|\tau)^{-1/6}\\
f_4(q) &= q^{-\frac{1}{24}}\prod_{m=1}^{\infty}(1-q^{2m-1})=
(2\pi)^{1/6}\theta_4(0|\tau)^{1/2}\theta'_1(0|\tau)^{-1/6}\nonumber\\ \nonumber \,.
\end{align}
where $q=e^{i \pi \tau}$.  Under the modular transformation $\tau \rightarrow -1/\tau$, $q \rightarrow \tilde q = e^{-i \frac{\pi}{\tau}}$ and   the functions transform as

\begin{align}
\label{eq:mtf}
f_1(\tilde q) & = \sqrt{-i\tau} f_1(q)\,, & 
f_2(\tilde q) & =  f_4(q)\nonumber\\
f_3(\tilde q) & = f_3(q)\,, &
f_4(\tilde q) & =  f_2(q)
\end{align}

\section{Density matrix for A system}
Following \cite{Gaberdiel:2000jr} we work with the notation $\alpha'=1$. 
In order to compute the density matrix, first we define the closed string hamiltonian in the NS-NS sector is 
\be
H_c = \pi {\bf P}^2 + 2\pi \sum_{\mu=2}^{9}\Big[\sum_{n=1}^{\infty}
\Big( \alpha^\mu_{-n} \alpha^\mu_n + {\tilde \alpha}^\mu_{-n}
{\tilde \alpha}^\mu_n
\Big)
+
\sum_{r=\frac{1}{2}}^{\infty} \Big( \psi^\mu_{-r} \psi^\mu_r + {\tilde \psi}^\mu_{-r}
 {\tilde \psi}^\mu_r
 \Big)   \Big] -2\pi
\ee
Then, for the NS-NS boundary state 
\be
\langle \langle \rho \rangle \rangle = \sum_{\eta \eta'} \eta \eta' e^{-\epsilon H_c}\ket{Bp,{\bf a},\eta}_{\NN}  {}_{\NN}\bra{Bp,{\bf a},\eta'} e^{-\epsilon H_c}
\ee

In order to compute $\rho_A$ we dissect the boundary state and the Hamiltonian in their  parts: the momentum contribution and the oscillator modes.

In the definition of $\rho_A$ given in \eqref{eq:densityA}, the trace in the momentum is denoted as $\Tr_C \rho$. 
The momentum contribution to the boundary state is expressed as a Fourier transform in the position space localizing the brane in  position ${\bf a}$ along the Dirichlet directions:
\be
\ket{B,{\bf a}} = \int \prod_{\n = 0,1,p+3}^9 d k^{\nu} e^{i k^{\nu}a_{\nu}}\ket{\bf k}\,.
\ee
Along the  Neumann directions the momentum is zero, so the boundary state carries momentum  ${\bf k}$  along the Dirichlet directions only.
To compute the trace we note that for a closed string  groud state with momentum ${\bf p}$

\be
\bra{{\bf p}} e^{-\pi \epsilon {\bf P}^2}\ket{B,{\bf a}} = e^{i p \cdot a}e^{-\pi \epsilon {\bf p}^2}\,.
\ee
Therefore  
\be
\label{eq:AtrC}
\Tr_C \langle \langle \rho \rangle \rangle= \Tr_C \Big[ e^{-\pi \epsilon {\bf P}^2} \ket{B,{\bf a}}\bra{B,{\bf a}} e^{-\pi \epsilon {\bf P}^2} \Big] = \Bigg(\frac{1}{2\epsilon} \Bigg)^{\frac{9-p}{2}}\,,
\ee

The trace over B, is the trace over the left-oscillator modes. We first consider the bosonic  oscillator mode in the definition of $\rho$. Taking the bosonic part of the string state defined in \eqref{eq:state} we have:

\be
\bra{l_L} e^{-\epsilon H_c}\ket{B}= \prod_{\mu=1}^{9} \prod_{n=1}^{\infty} e^{-4 \pi \epsilon n l_n } \frac{(-1)^{l_n}}{\sqrt{l_n}!} \Bigg(\frac{\alpha^\m_{-n}}{\sqrt{n}}  \Bigg)^{l_n}\ket{0}
\ee
For the fermionic part, we have to consider the contribution to the density matrix of the boundary state with spin structure,that is:
\be
\bra{l_L} e^{-\pi \epsilon H_c}\ket{B,\eta} = \prod_{\mu=1}^{9} \prod_{r=\frac{1}{2}}^{\infty} \Big[ \delta_{l^\mu_r,0} + i\eta e^{-4\pi \epsilon r}S_{\mu \mu}\delta_{l^\m_r,1} \psi^\m_{-r}\Big] \ket{0} 
\ee
Defining $q=e^{-4 \pi \epsilon}$ we have

\be
\label{eq:AtrB}
\Tr_B \langle \langle \rho \rangle \rangle = q^{-1}\prod_{\mu} \Bigg[\prod_{n}\sum_{l^\m_n}  q^{2 n l^\m_n}
\frac{1}{l^\m_n!} \Big(\frac{\alpha^\m_{-n}}{\sqrt{n}}  \Big)^{l^\m_n}\otimes \Big(\frac{\alpha^\m_{n}}{\sqrt{n}}  \Big)^{l^\m_n} \Bigg]\Bigg[ \prod_r(1 \otimes 1 + \eta \eta'q^{2r}\psi^\mu_{-r}\otimes{\psi^\mu_r}) \Bigg] \ket{0}\otimes \bra{0}
\ee
where we have included in the bosonic part the contrubution of the zero-point energy. The notation for the tersor product is that for $A$ y $B$ operators, $(A \otimes B)(\ket{0}\otimes\bra{0}) = A\ket{0} \otimes \bra{0}B$. Finally,the desired trace is

\be
\label{eq:trpA}
\begin{split}
 \langle \langle \rho_A \rangle \rangle = \langle \langle \Tr_B(\Tr_C \rho )\rangle \rangle  &={\cal N}^2  \Bigg(\frac{1}{2\epsilon} \Bigg)^{\frac{9-p}{2}} q^{-1} 
 \prod_{\mu} \prod_{n}\sum_{l^\m_n}  q^{2 n l^\m_n}
\frac{1}{l^\m_n!} \Big(\frac{\alpha^\m_{-n}}{\sqrt{n}}  \Big)^{l^\m_n}\otimes \Big(\frac{\alpha^\m_{n}}{\sqrt{n}}  \Big)^{l^\m_n} \\
&\times
 \sum_{\eta \eta} \eta \eta'
\Bigg[\prod_{\mu} \prod_r(1 \otimes 1 + \eta \eta'q^{2r}\psi^\mu_{-r}\otimes{\psi^\mu_r}) \Bigg] \ket{0}\otimes \bra{0}
\end{split}
\ee

For $\rho_{AC}$ we do not sum over the momentum, then we have:

\be
\label{eq:trac}
\begin{split}
\langle \langle \rho_{AC} \rangle \rangle &= \Tr_B   \langle \langle \rho_{AC} \rangle \rangle   = {\cal N}^2 \int \prod_{\nu} \int \prod_{\nu'} dk^{\nu} dk'^{\nu'}e^{i{\bf a}\cdot ({\bf k} - {\bf k}')}e^{-\pi ({\bf k}^2 - {\bf k}'^2)}q^{-1} \times \\
&\prod_{\mu} \Bigg[\prod_{n}\sum_{l^\m_n}  q^{2 n l^\m_n}
\frac{1}{l^\m_n!} \Big(\frac{\alpha^\m_{-n}}{\sqrt{n}}  \Big)^{l^\m_n}\otimes \Big(\frac{\alpha^\m_{n}}{\sqrt{n}}  \Big)^{l^\m_n} \Bigg]\Bigg[ \prod_r(1 \otimes 1 + \eta \eta'q^{2r}\psi^\mu_{-r}\otimes{\psi^\mu_r}) \Bigg] \ket{0,{\bf k}}\otimes \bra{0,{\bf k}}
\end{split}
\ee

\section{Supersymmetric branes do not get entangled!}
This appendix has a two-fold motivation. First we want to display explicit aspects of the calculation of the entanglement entropy in the context of D-brane and second we want to highlight the fact that for supersymmetric branes the entanglement vanishes identically when sum over all spin structures.

The vanishing of the entanglement is akin to the vanishing one-loop diagram in string theory between two D-branes and we learn that to get a non-trivial entanglement one needs to consider sectors of a given spin structure.
	
\bibliographystyle{JHEP}
\bibliography{entanglement}

\end{document}